\documentclass[journal,twoside,web]{ieeecolor}
\usepackage{tmi}
\usepackage{cite}
\usepackage{url}
\usepackage{amsmath,amssymb,amsfonts}
\usepackage{algorithmic}
\usepackage{mathtools} 
\usepackage{multirow}
\usepackage{graphicx}

\usepackage{overpic}
\usepackage{textcomp}
\usepackage{floatrow}
\usepackage{arydshln} 
\usepackage{url}
\usepackage{caption}
\usepackage{xcolor}
\usepackage{bbding}
\usepackage{hyperref}
\usepackage{graphicx}
\usepackage[misc]{ifsym}
\usepackage{pgfplots}
\usepackage{tikz}
\usepackage[ruled]{algorithm2e}
\usepackage{multirow}
\usepackage{xr}
\usepackage{soul} 
\usepackage{graphicx} 
\usepackage{float} 
\usepackage{subfigure}  
\usepackage{sidecap}
\usepackage{scrextend}
\usepackage{threeparttable}
\usepackage[switch]{lineno}

\newcommand{\hly}[1]{{#1}}
\newfloatcommand{capbtabbox}{table}[][\FBwidth]

\SetCommentSty{mycommfont}
\usepackage{bm}

\newcommand{\ie}{\textit{i.e.}, }
\newcommand{\MyFig}{Fig. }
\newcommand{\MyTab}{Table. }
\newcommand{\eg}{e.g. }

\makeatletter
\newcommand*{\addFileDependency}[1]{
  \typeout{(#1)}
  \@addtofilelist{#1}
  \IfFileExists{#1}{}{\typeout{No file #1.}}
}
\makeatother

\newcommand*{\myexternaldocument}[1]{
    \externaldocument{#1}
    \addFileDependency{#1.tex}
    \addFileDependency{#1.aux}
}

\myexternaldocument{supply}
\definecolor{newcolor}{rgb}{.8,.349,.1}

\def\BibTeX{{\rm B\kern-.05em{\sc i\kern-.025em b}\kern-.08em
    T\kern-.1667em\lower.7ex\hbox{E}\kern-.125emX}}
\begin{document}

\title{ 
CineMyoPS: Segmenting Myocardial Pathologies from Cine Cardiac MR}
\author{
%First A. Author, \IEEEmembership{Fellow, IEEE}, Second B. Author, and Third C. Author, Jr., \IEEEmembership{Member, IEEE}
 Wangbin  Ding, Lei Li, Junyi Qiu, Bogen Lin,  Mingjing Yang, Liqin Huang, 
 %Yinyin Chen, Shan Yang, 
 Lianming Wu, Sihan Wang,  Xiahai Zhuang
%Wangbin Ding, Xiahai Zhuang
\thanks{%M. Yang and X. Zhuang are co-senior authors and contribute equally. 
Corresponding authors: M. Yang, L. Huang, L. Wu and X. Zhuang. 
This work was supported by the National Natural Science Foundation of China (62401148, 62372115, 62271149); the Shanghai Municipal Education Commission-Artificial Intelligence Initiative to Promote Research Paradigm Reform and Empower Disciplinary Advancement Plan (24KXZNA13); the Research Initiation Project of Fujian Medical University (XRCZX2024003); the Science and Technology Project of Fujian Province (2020Y9037, 2021Y9220, 2020Y9091, 2022Y4014, 2024J01353).
} 
%\thanks{Corresponding authors: X. Zhuang. This work was supported by the National Nature Science Foundation of China, 62111530195, 61971142 and 62271149; Fujian Provincial Natural Science Foundation Project 2021J02019. } 
%\thanks{M. Yang and X. Zhuang are co-senior authors and contribute equally. Corresponding authors: M. Yang, X. Zhuang, Y. Chen and S. Yang.  This work was supported by the National Nature Science Foundation of China, 62111530195, 61971142 and 62271149; Fujian Provincial Natural Science Foundation Project 2021J02019. } 
%\thanks{M. Yang and L. Huang are with the College of Physics and Information Engineering, Fuzhou University, Fuzhou 350117, China (n191110003@fzu.edu.cn; hlq@fzu.edu.cn).}
\thanks{W. Ding is with the School of Medical Imaging, Fujian Medical University, Fuzhou 350117, China. (dingwangbin@fjmu.edu.cn)}
\thanks{B. Lin, L. Huang and M. Yang  are with the College of Physics and Information Engineering, Fuzhou University, Fuzhou 350117, China ({221127167@fzu.edu.cn}; hlq@fzu.edu.cn; yangmj5@fzu.edu.cn).} 
% \thanks{L. Li is with the School of Electronics \& Computer Science, University of Southampton, SO17 1BJ Southampton, UK and Department of Engineering Science, University of Oxford, OX3 7DQ Oxford, UK (lei.sky.li@soton.ac.uk).}
\thanks{L. Li is with the Department of Biomedical Engineering, National University of Singapore, Singapore (lei.li@nus.edu.sg).}
\thanks{J. Qiu, S. Wang and X. Zhuang are with the School of Data Science, Fudan University, Shanghai, China (qjy980811@gmail.com; 21110980009@m.fudan.edu.cn; zxh@fudan.edu.cn).}
%\thanks{X. Zhuang is with the School of Data Science, Fudan University, Shanghai, China ( zxh@fudan.edu.cn).}
%\thanks{Y. Chen and S. Yang are with the Department of Radiology, Zhongshan Hospital, Fudan University and also the Department of Medical Imaging, Shanghai Medical School, Fudan University and Shanghai Institute of Medical Imaging, Shanghai 200032, China (chen.yinyin@zs-hospital.sh.cn; yang.shan@zs-hospital.sh.cn).}
\thanks{L. Wu is with the Department of Radiology, Renji Hospital, School of Medicine, Shanghai Jiao Tong University, Shanghai, China (wlmssmu@126.com)}
}
\maketitle

\begin{abstract}

Myocardial infarction (MI) is a leading cause of death worldwide. Late gadolinium enhancement (LGE) and T2-weighted cardiac magnetic resonance (CMR) \hly{imaging} can respectively identify scarring and edema areas, \hly{both of which} \hly{are} essential for MI risk stratification and prognosis assessment. Although combining complementary information from multi-sequence CMR is useful, acquiring these sequences can be time-consuming and prohibitive, e.g., due to the administration of contrast agents. Cine CMR is a rapid and contrast-free imaging technique that can visualize both motion and structural \hly{abnormalities} of the myocardium induced by acute MI. Therefore, we present a new end-to-end deep neural network, referred to as CineMyoPS, to segment myocardial pathologies, \ie scars and edema,  solely from cine CMR images. Specifically, CineMyoPS extracts both motion and anatomy features \hly{associated with MI}. \hly{Given} the interdependence \hly{between these features}, we design a consistency loss (resembling the co-training strategy) to \hly{facilitate their joint learning}. Furthermore, we propose a time-series aggregation strategy to integrate \hly{MI-related} features across the cardiac cycle, \hly{thereby enhancing segmentation accuracy for} myocardial pathologies. \hly{Experimental} results on a multi-center dataset demonstrate that CineMyoPS achieves promising performance in myocardial pathology segmentation, motion estimation, and anatomy segmentation.

\end{abstract}


\deffootnote[2em]{2em}{1em}{\textsuperscript{\thefootnotemark}\,}

\section{Introduction}\label{sec:intro}

The incidence of myocardial infarction (MI) is rising worldwide \cite{chandrashekhar2020resource}. MI is \hly{typically} caused by blockages in the coronary arteries, leading to the formation of \hly{the} area at risk (AAR) in \hly{the} myocardium. In acute MI cases, the AAR typically consists of scarring and edema regions \cite{dall2011dynamic}. Clinicians can calculate the myocardial salvage index and assess the prognosis of MI patients based on these two regions~\cite{beijnink2021cardiac}. Thus, AAR quantification is critical for clinical applications.

\begin{figure*}[htpb]
    \centering
    \includegraphics[width=\textwidth]{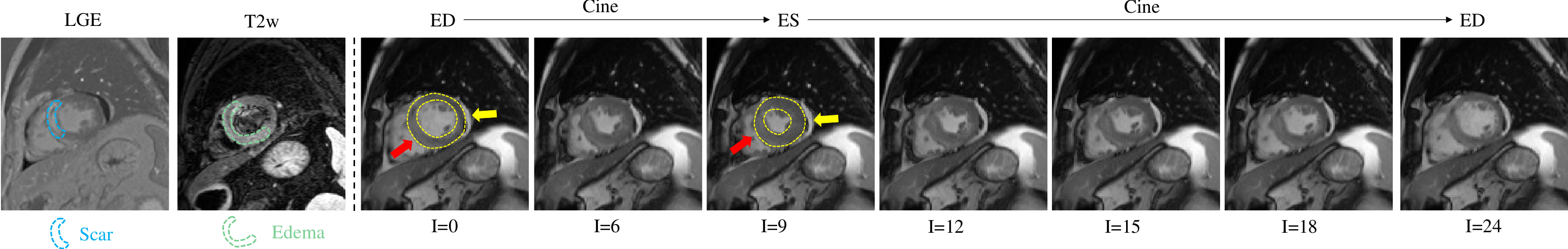}
    \caption{Representative LGE, T2w, and cine CMR \hly{images from an MI patient}.  The infarcted myocardium remains relatively \hly{stationary} (indicated by red arrows), whereas the viable myocardium \hly{exhibits noticeable deformation} (indicated by yellow arrows). This \hly{observation suggests} that myocardial pathologies can induce abnormal motion patterns in the myocardium. 
    ED: end-diastolic frame, ES: end-systolic frame.
    }
    \label{fig:intro}
    
\end{figure*}

Cardiac magnetic resonance (CMR) imaging techniques can visualize myocardial structure, motion, and pathological \hly{changes} \cite{li2023multi}, \hly{making them invaluable for} quantifying the AAR in MI patients \cite{li2023myops}. As shown in the left \hly{panel} of \MyFig \ref{fig:intro}, late gadolinium enhancement (LGE) CMR visualizes scarring areas, while T2-weighted (T2w) CMR \hly{identifies}  edema \cite{li2023myops}. However,  these techniques face clinical limitations. For instance, the average acquisition time of {multi-sequence CMR (MS-CMR)}, comprising T2w, first-pass perfusion, cine, and LGE, is approximately 32$\pm$8 minutes \cite{cury2008cardiac}, making it impractical for severe acute MI \hly{cases}. In particular,  acquiring \hly{the} LGE sequence is time-consuming, \hly{accounting for approximately 50\% of the total examination duration \cite{polacin2021segmental}}. Moreover, the potential harm of gadolinium deposition remains under debate \cite{gulani2017gadolinium}, and there is a risk of cross-reactivity with other macrocyclic contrast agents \cite{vega2024cross}. These limitations highlight the need for \hly{alternative methods that require} fewer CMR sequences and avoid the use of contrast agents \cite{bulluck2017quantification}.     

 MI can \hly{induce} abnormal motion patterns in the myocardium, and the \hly{analysis } of these patterns has been widely investigated for infarction localization.  Medical \hly{imaging modalities}, such as cine CMR images, \hly{are} capable of capturing these motion abnormalities as well as structural {malformations}, as shown \hly{in} the right \hly{panel} of \MyFig \ref{fig:intro}. Various methods have been proposed \hly{to localize} infarcts based on myocardial deformation patterns, \hly{with the aim of reducing} \hly{the} risks associated with contrast agents. One of the most straightforward approaches \hly{is} to threshold \hly{the} deformation patterns \cite{sjoli2009diagnostic}. To \hly{achieve} better localization accuracy, more advanced machine learning techniques have been employed, such as kernel regression \cite{duchateau2016infarct}, \hly{dictionary learning \cite{peressutti2015towards}}, and random forests \cite{bleton2016myocardial}. \hly{In addition}, due to the complexity of deformation patterns,  strain parameters \cite{rumindo2017strain, mangion2017magnetic} and embedding techniques \cite{duchateau2015prediction,duchateau2012constrained} have been incorporated to improve scar identification performance.

Recently, deep neural network (DNN) techniques have improved non-contrast quantification of myocardial pathology \hly{by leveraging} cine CMR images \cite{xu2020segmentation,zhang2019deep,xu2020contrast}. 
These methods can be categorized into three groups:
\begin{itemize}
    \item Motion estimation-based methods \cite{zhang2019deep}: These methods establish associations between motion fields and scars \hly{using} DNN. \hly{However}, they \hly{rely on} an external optical flow algorithm \cite{chen2016full} to extract wall motion from cine CMR images. 
    
    \item Spatio-temporal convolution network-based methods \cite{xu2020segmentation}: Since cine CMR images \hly{capture both} spatial and temporal information of \hly{the} myocardium, these methods \hly{employ} convolution layers, such as ConvLSTM \cite{shi2015convolutional} and 3D Unet \cite{tran2015learning}, to \hly{extract} underlying spatio-temporal features for scar prediction.

    \item Synthesis-contrast-based methods \cite{xu2020contrast}: These methods utilize conditional generative adversarial \hly{networks} (GANs) \cite{goodfellow2020generative}  to synthesize LGE-\hly{equivalent} images, \hly{from which} scars \hly{can} be directly extracted using existing segmentation methods, such as \hly{Unet} \cite{ronneberger2015u}.
\end{itemize}

\hly{Even though} cine CMR provides a fast and contrast-free way for \hly{myocardial pathology quantification}, existing cine CMR-based methods \hly{primarily} focus on scar segmentation. \emph{Limited attention has been paid to edema pathology}, which is critical in identifying the potential rescued heart tissue after an intervention \cite{o2013salvage}. 
Nevertheless, several \hly{studies have demonstrated} that cine CMR \hly{sequences exhibit} internal consistency in the assessment of edema in \hly{the} myocardium, \hly{providing} a new opportunity to jointly assess myocardial scars and edema from \hly{the} cine CMR \cite{kumar2008t2,goldfarb2011cyclic}. 

In this work, we propose an end-to-end \hly{myocardial pathology segmentation (MyoPS)} network, referred to as CineMyoPS, based on DNN techniques. To \hly{the best of our} knowledge, CineMyoPS is the first fully automatic network for joint scars and edema segmentation from cine CMR images. Specifically, it introduces a motion estimation module to track myocardial movements and a segmentation module to extract anatomical structures from cine CMR images. Given that both modules \hly{rely} on structural features, we introduce a consistency loss to improve the feature extraction of the two modules. To further \hly{enhance} MyoPS performance, we propose a time-series aggregation strategy \hly{that flexibly fuses} myocardial motion, anatomy, and texture features \hly{across} cardiac cycles.  
structure
Overall, CineMyoPS is an extension of motion estimation-based methods, as it captures various features, including motion, \hly{to enhance} MyoPS. Additionally, it is inspired by spatio-temporal methods, specifically \hly{leveraging} time-series information to enhance MyoPS across cardiac cycles.

\section{Method}

\begin{figure*}
    \centering
    \includegraphics[width=1\textwidth]{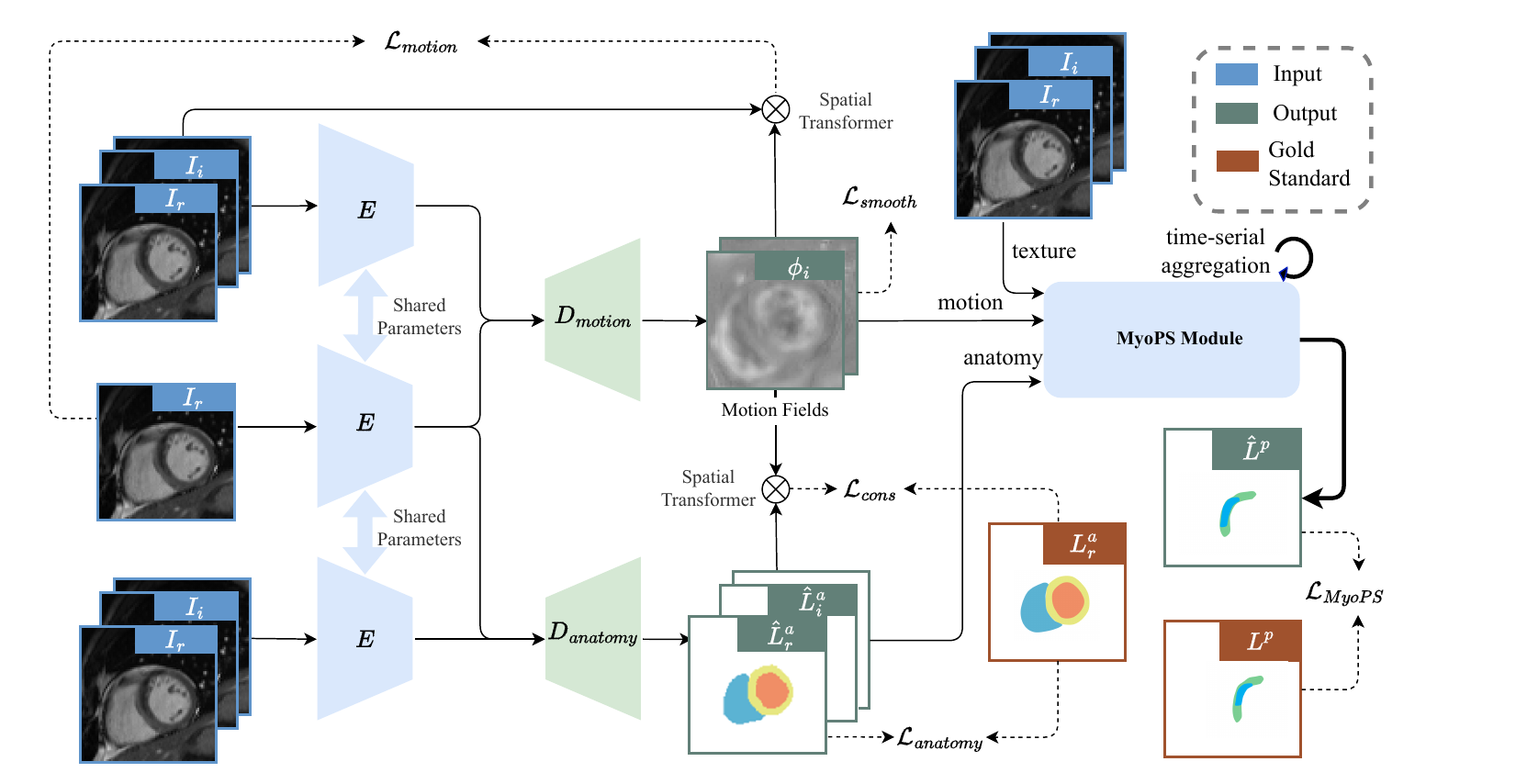}
    \caption{The architecture of the myocardial pathology segmentation (MyoPS) network for cine CMR \hly{images}, \ie CineMyoPS. It consists of three modules:  a motion estimation module (consisting of $E$ and $D_{motion}$),  an anatomy segmentation module (consisting of $E$ and $D_{anatomy}$), and a MyoPS module. The motion estimation module extracts motion \hly{from the} cine CMR images, while the anatomy segmentation module predicts myocardial structures \hly{from the same images}. \hly{The} MyoPS module aggregates motion, anatomy, and texture features (represented by \hly{the} original cine CMR images) for pathology segmentation. For details of \hly{the} MyoPS module, please refer to \MyFig \ref{fig:myops_net}.}
    \label{fig:framework}
\end{figure*}

Fig. \ref{fig:framework} shows the network architecture of CineMyoPS, 
which consists of three modules: a myocardial motion estimation module (see Section \ref{sec:motion}), an anatomy segmentation module (see Section \ref{sec:myoseg}), and a MyoPS module (see Section \ref{sec:myops}). CineMyoPS aggregates multiple types of features along a cardiac cycle and predicts myocardial pathologies, \hly{including} scars and edema, \hly{in a reference image space.} 

\subsection{Myocardial Motion Estimation}
\label{sec:motion}

The extent of wall-motion abnormality \hly{is strongly correlated with} the AAR and infarction size  \hly{in MI patients} \cite{rodrigues2004relationship}. We introduce a motion estimation module to capture motion \hly{from} cine CMR sequences. Let $\mathcal{\mathbf{I}}=\{I_i| i=1\cdots n\}$ be a cine CMR sequence with $n$ frames.  We set the end-diastolic (ED) frame of $\mathcal{\mathbf{I}}$ as \hly{a common reference image} (\ie $I_{r}$), and introduce a network to estimate \hly{the} motion between $I_{r}$ and \hly{each} $I_i$.

\begin{figure}
    \centering
    \includegraphics[width=\textwidth]{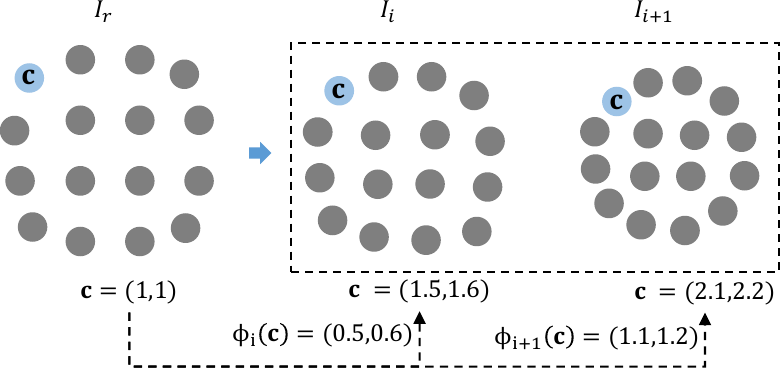}
    \caption{A representative example of the motion field in a cine CMR sequence. Each element of the motion field (\eg  $\Phi_{i}(\textbf{c})$) represents the \hly{displacement} of $\textbf{c}$, where $(x,y)$ \hly{denotes} its  coordinates.}
    \label{fig:motion_intro}
\end{figure}

The backbone of the motion estimation module is a U-shaped registration subnetwork {\cite{balakrishnan2019voxelmorph}}. It \hly{takes} $I_i$ and $I_{r}$ \hly{as input} and predicts \hly{a} dense displacement field (DDF) between them. Formally, the motion estimation module \hly{can be} defined as follows:
\begin{equation}
    {\Phi}_{ i}=\mathcal{F}_{motion}(I_{r},I_i),
\end{equation}
where $\Phi_{ i}$ is the predicted DDF. We \hly{can transform} $I_i$  to $I_{r}$ by using the corresponding DDF as follows:
\begin{equation}
    \tilde{I}_{i}=I_i\otimes \Phi_i,
\end{equation}
where $\otimes$ \hly{denotes} the image transform operation, and each element in $\tilde{I}_i$ is calculated as follows:
\begin{equation}
   \tilde{I}_{i}(\textbf{c})=I_i(\textbf{c}+ \Phi_{ i}(\textbf{c})),
\end{equation}
where $\textbf{c}$ denotes a spatial point  and $\Phi_{ i}(\textbf{c})$ \hly{represents} the corresponding displacement at $\textbf{c}$. 
Note that $\Phi_{i}$ is defined within \hly{the reference image space}.  Each element of $\Phi_{i}$, \eg  $\Phi_{i}(\textbf{c})$, represents the movement of \textbf{c} from the time point of $I_r$ to $I_i$, as shown in \MyFig \ref{fig:motion_intro}. 
By estimating the DDF between $I_i$ and $I_r$, we obtain the motion pattern of $\textbf{c}$, which \hly{can} further be applied to MyoPS. 

The motion estimation module can be trained by minimizing the difference between the reference image $I_r$ and the moved image $\tilde{I}_i$:
\begin{equation}
    \mathcal{L}_{motion}=\sum_{i=1}^{n}\text{MSE}(\tilde{I}_i, I_r ),
\end{equation}
where $\text{MSE}(a,b)$ \hly{computes} the mean square error between $a$ and $b$. Meanwhile, we constrain the smoothness of  $\Phi_i$ \hly{through}  a regularization term:
\begin{equation}
    \mathcal{L}_{smooth}=\sum_{i=1}^{n}\| \nabla \Phi_{i}  \|_2^2,
\end{equation}

\subsection{Anatomy Segmentation and Consistency Loss}
\label{sec:myoseg}
MI often induces geometric remodeling of the left ventricle, \hly{leading to myocardial wall thinning} \cite{d2015association}.  
We introduce an anatomy segmentation module to extract myocardial structures \hly{from the} cine CMR sequences. 
The backbone of the module is a U-shaped segmentation subnetwork {\cite{ronneberger2015u}}. It \hly{takes}  $I_i$ as input and predicts segmentation as follows:
\begin{equation}
    \hat{L}^{a}_{i}=\mathcal{F}_{anatomy}(I_{i}), 
\end{equation}
where $\hat{L}^{a}_i$ denotes the predicted  segmentation result of $I_i$. 
The parameters of the anatomy segmentation module \hly{can} be trained by minimizing the difference between the predicted and gold standard segmentation of $I_i$. However, delineating all frames in a cine CMR sequence is time-consuming, as \hly{each sequence} typically \hly{contains} 25 to 30 frames. To address this, we delineate only the \hly{anatomical} label of the ED frames, and \hly{introduce} a consistency loss to \hly{regularize the segmentation results across the cardiac cycle}.

\hly{Specifically}, we \hly{first train} the anatomy segmentation module using  a supervised loss: 
\begin{equation}
    \mathcal{L}_{anatomy}=-\text{Dice}(\hat{L}^{a}_{r},{L}^{a}_{r})-\text{CE}(\hat{L}^{a}_{r},{L}^{a}_{r}),
\end{equation}
where $\text{Dice}(a,b)$ and $\text{CE}(a,b)$ \hly{compute} the Dice score and cross-entropy between $a$ and $b$, respectively. \hly{$L^a_{r}$} is the gold standard label of $I_r$.
Then, given an unlabeled image $I_i$,  we \hly{assume} the anatomy of $I_i$ (\ie $\hat{L}^{a}_{i}$), should be consistent \hly{with that of $I_r$ (i.e. $L^a_r$)}, after transforming \hly{it} with $\Phi_{i}$. We follow the joint segmentation and motion estimation schema \hly{previously proposed in the} literature \cite{qin2018joint,ta2020semi}, and further explore \hly{a} consistency loss based on cosine distance as follows: 
\begin{equation}
    \mathcal{L}_{cons}=1-\cos (\hat{L}^{a}_{i}\otimes\Phi_{ i} ,\hat{L}^{a}_r) , 
    \label{eq:cotrain}
\end{equation}
where $\cos(a,b)$ \hly{measures} the cosine \hly{similarity} \hly{between} $a$ and $b$.  Finally, $\mathcal{L}_{cons}$ and $\mathcal{L}_{anatomy}$ \hly{are jointly leveraged to train} the motion estimation and anatomy segmentation modules.

\subsection{Myocardial Pathology Segmentation}
\label{sec:myops}
Motion and anatomy features are strongly associated with myocardial pathologies, as MI \hly{often induces} abnormal motion \hly{patterns} and \hly{thinning} of the myocardial wall. 
Meanwhile, texture features \hly{derived from} the cine CMR sequences exhibit T1- and T2-\hly{weighted characteristics}, which may contribute to the performance of MyoPS \cite{tahir2017acute,jaubert2020free}.
We introduce a MyoPS module that \hly{integrates these} features to segment scarring and edema areas. In addition, the module is designed to effectively fuse time-series information from cine CMR images to enhance segmentation performance.

\begin{figure}[htpb]
    \centering
    \includegraphics[width=0.8\linewidth]{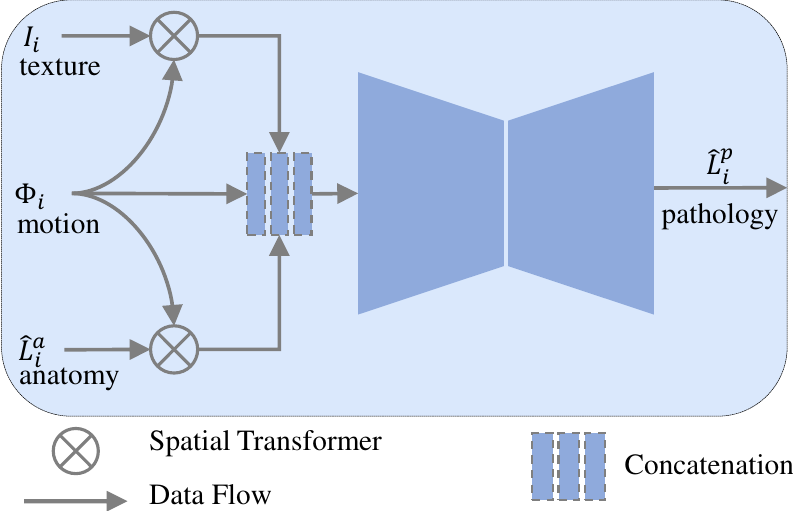}
    \caption{The segmentation network architecture of the MyoPS module. It \hly{takes various types of features as inputs,} including motion, texture, and anatomy,  to perform  MyoPS in a reference image space. }
    \label{fig:myops_net}
\end{figure}

\MyFig \ref{fig:myops_net} shows the architecture of the MyoPS module, which is a U-shaped segmentation subnetwork {\cite{ronneberger2015u}}. For each \hly{set of features} extracted from the $i$-th frame of $\mathcal{\mathbf{I}}$, \ie $\Phi_i, \hat{L}^{a}_i$ and $I_i$, the MyoPS module takes them as inputs and \hly{predicts}  myocardial pathologies \hly{in the} reference image space as follows: 
\begin{equation}
    \hat{L}^{p}_i=\mathcal{F}_{MyoPS}([{\Phi}_i, \overset{\text {}}{\hat{L}^{a}_i\otimes\Phi_{ i}}, \overset{\text {}}{I_i \otimes\Phi_{ i}}]),
\end{equation}
where $ \hat{L}^{p}_i$ denotes the predicted MyoPS result of the $i$-th frame, $[\cdots]$ \hly{represents the} concatenation operation. 
Note that segmentation is performed in the reference image space. The MyoPS module transforms \hly{the} anatomy ($\hat{L}^{a}_i$) and texture $I_i$  by \hly{the} motion field ($\Phi_i$) to mitigate spatial misalignment with the reference image $I_r$.

Furthermore, the MyoPS module can aggregate time-series information for segmentation. Given $n$ \hly{sets of} $\Phi_i$, $\hat{L}^{a}_i$, and $I_i$, the module \hly{generates} $n$ potential results across the  cardiac cycle. These time-series results \hly{are subsequently integrated} to obtain the final segmentation as follows:
\begin{equation}
    \hat{L}^{p}=\text{Softmax}(\text{conv}(\sum_i^n \hat{L}^{p}_i)),
    \label{eq:fusing}
\end{equation}
where $ \hat{L}^{p}$ denotes the final MyoPS result, and $\text{conv}(\cdots)$ represents a convolution operation.

Having $\hat{L}^{p}$ estimated, the network \hly{parameters} of the MyoPS module \hly{can} be trained as follows:
\begin{equation}
    \mathcal{L}_{MyoPS}=-\text{Dice}(\hat{L}^{p},{L}^{p})-\text{CE}(\hat{L}^{p},{L}^{p}),
        \label{eq:myops}
\end{equation}
where ${L}^{p}$ denotes the gold standard myocardial pathology label of \hly{the reference image $I_r$}. 

Finally, we jointly train the myocardial motion estimation, anatomy segmentation, and MyoPS modules in an end-to-end \hly{manner}. The overall loss function of CineMyoPS is as follows:
\begin{equation}
\begin{split}
    \mathcal{L}=&\mathcal{L}_{MyoPS}+\lambda_1 \mathcal{L}_{anatomy}+\lambda_2 \mathcal{L}_{cons}\\
    &+\lambda_3 \mathcal{L}_{motion}+\lambda_4 \mathcal{L}_{smooth},
\end{split}
\label{eq:totalloss}
\end{equation}
where $\lambda_1$, $\lambda_2$, $\lambda_3$, and $\lambda_4$ are the hyperparameters.  

\section{Experiments}
\label{sec:exp}
 This section first introduces the experimental datasets, implementation details, and evaluation metrics. Next, we conduct parameter and ablation studies for frame interval selection, feature effectiveness investigation, and consistency loss evaluation. Following this, we compare CineMyoPS to state-of-the-art methods. Finally, we assess the clinical quantification capabilities of our method.

\subsection{Datasets}

We evaluated CineMyoPS on a multi-center dataset, which includes three centers referred to as Center-M, Center-R, and Center-Z. The dataset contains MS-CMR sequences (\ie cine, LGE, and T2w) from 145 MI patients. 
Tables \ref{tab:demograph} and Table \ref{tab:param:cmr} provide the demographic information of the patients and the acquisition parameters of the CMR sequences, respectively.  \hly{Each case was independently annotated by two experienced raters. Annotations agreed upon by both raters were accepted as the gold standard, while disagreements were resolved by a senior radiologist specializing in cardiac imaging.}

\begin{table}[htp]
	\centering
	\caption{Demographic information of Center-M, Center-Z, and Center-R. PRE and POST refer to whether the cine images were acquired pre-contrast or post-contrast.  SBP: systolic blood pressure; DBP: diastolic blood pressure; HR: heart rate; TR: training; VA: validation; TE: testing. Num: number. Male values are reported as the ``number of male cases (percentage\%)'' format. Age, weight, SBP, DBP, and HR values are presented as the ``median [min, max]'' format. NA: Not available.}
%\resizebox{0.9\textwidth}{!}{
\begin{tabular}{lccc}
\hline
 & Center-M    & Center-Z   & Center-R \\
 \hline
Num & 45 & 50 &50 \\
{Contrast} &  {PRE \& POST} & {POST} & {PRE} \\
{Male}           & {45 (100\%)}     & {48 (96\%) }   &    40 (80\%)      \\

Age             & {57 [38, 69]} & {56 [28, 69]} &   60 [36,77]   \\
{Weight (kg)} & {75 [60, 90]} & {71 [54,102]}  & 72 [50,90] \\
{SBP(mmHg)}&  NA & {120 [92,186]}&  NA \\
{DBP(mmHg)}&  NA & {76 [48,113]}&  NA \\
{HR} &  NA & {78 [55,114]} &  NA \\
TR/VA/TE & {35/10/0}      & {40/10/0}      &    {0/0/45}     \\

\hline
\end{tabular}
\label{tab:demograph}
\end{table}

\begin{table*}[htp]
    \centering
   \resizebox{0.6\textwidth}{!}{ 
    \begin{tabular}{lcccc}
    \hline
        Seqs &  In-plane resolution (mm$^2$) & SPC (mm)  &Slice Num & Frame Num \\
        \hline
           Center-M\\ 
        
          \hline
        Cine &  1.19 $\times$ 1.19 $\sim$ 1.25 $\times$ 1.25  & 12 $\sim$ 23   & 2 $\sim$ 6 & 30 \\
               
        T2w &  0.73 $\times$ 0.73  $\sim$ 0.76 $\times$ 0.76  & 12 $\sim$ 20   & 2 $\sim$ 6   & 1 \\
            LGE       &  0.73 $\times$ 0.73  $\sim$ 0.76 $\times$ 0.76  & 11 $\sim$ 23   & 2 $\sim$ 6  & 1 \\

          \hline
         Center-R\\
                \hline
                Cine & 1.17 $\times$1.17  & 6  &  5 $\sim$ 15 & 30 \\
        
         LGE &  0.89 $\times$ 0.89   &  10   &  3 $\sim$ 12  & 1 \\
                
        T2w &  0.89 $\times$ 0.89  & 10 $\sim$ 17   &  3 $\sim$ 11 & 1  \\
          \hline
         Center-Z \\ 
         \hline
        Cine &  1.77 $\times$ 1.77 $\sim$ 2.13 $\times$ 2.13  & 10   & 2 $ \sim $ 7 & 25 \\
        
        LGE &  1.33 $\times$ 1.33 $\sim$ 1.86 $\times$ 1.86   & 10   & 4 $ \sim $ 11 & 1\\
                
        T2w &  1.33 $\times$ 1.33 $\sim$ 2.24 $\times$ 2.24   & 10   & 2 $\sim $ 10 & 1 \\
          \hline
    \end{tabular}
    }
    \caption{Acquisition parameters of multi-sequence CMR. Seqs: sequence, SPC: slice spacing (thickness + gap).}
    \label{tab:param:cmr}
\end{table*}

\begin{figure}[htp]
    \centering
    \includegraphics[width=0.8\linewidth]{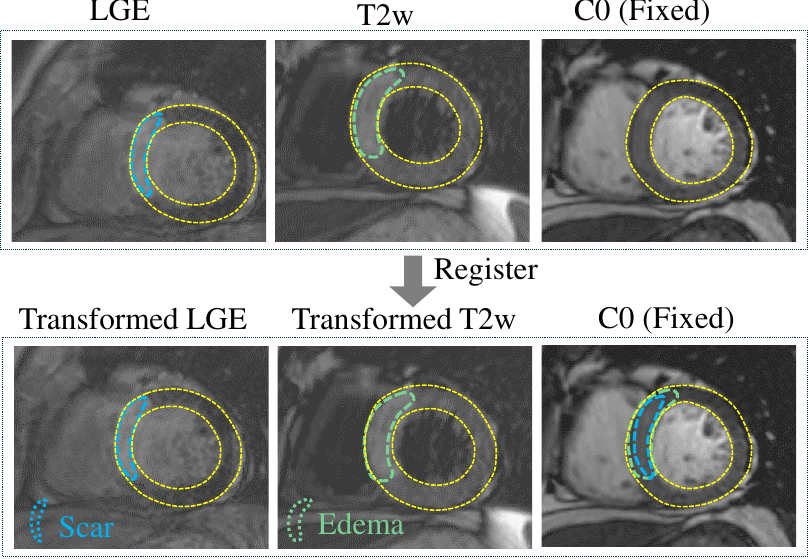}
    \caption{\hly{Illustration of the} gold standard myocardial pathology label \hly{generation process}. C0: End-diastolic frame of the cine CMR sequence.}
    \label{fig:myops_gen}
\end{figure}

The gold standard pathologies (\hly{$L^p$}) were generated by fusing the pathology labels from LGE and T2w images. 
As shown in \MyFig \ref{fig:myops_gen}, we first annotated scar and edema labels in LGE and T2w images, respectively.  Then, we registered the two images to the ED phase of the cine CMR sequence from the same subject using the MvMM tool \cite{zhuang2018multivariate}.  \hly{In this step, structural information from all three image modalities was utilized to ensure} accurate registration. Finally, the \hly{registered} labels were fused to \hly{produce} the gold standard. 

In \hly{the} experiments, Center-M and Center-Z were randomly divided into a training set of 75 subjects and a validation set of 20 subjects, while Center-R was used for testing. The parameter and ablation studies (see Section \ref{sec:Parameter and Ablation Study}) \hly{were} conducted on the validation dataset. The comparative and clinical quantification studies (see Section \ref{sec:Performance and Comparisons} and Section \ref{sec:clnical_study}) were \hly{performed} on the testing dataset.

\subsection{Implementation and Evaluation Metrics}

CineMyoPS was implemented in PyTorch and trained using an RTX 4090 GPU\footnote{Source code will be released publicly at \url{https://github.com/NanYoMy/CineMyoPS} once the manuscript is accepted.}. The loss function of CineMyoPS consists of four hyperparameters [see Equation \eqref{eq:totalloss}]. For each hyperparameter, we began \hly{with} a coarse search over predefined values (0.01, 0.1, 1, 10, 100).  The parameter value was initially identified based on the Dice score of scar and edema segmentation in the validation set. \hly{Subsequently, a fine-grained linear search was conducted within a range,} from the identified value to five times the identified value, to determine the optimal one. The final hyperparameters of the overall loss function \eqref{eq:totalloss}  $\lambda_1$,  $\lambda_2$,  $\lambda_3$,  and $\lambda_4$ were set to 5, 2, 1, and 100, respectively.

To evaluate \hly{the performance of CineMyoPS}, the Dice score and Hausdorff distance (HD)  were calculated between the predicted results and the gold standards. Additionally, sensitivity (Sen), precision (Pre), \hly{specificity (Spe), and negative predictive value (NPV)} were included to assess the effectiveness of the method. \hly{Note that we referred to the myocardium as the negative class when calculating Spe and NPV.} Furthermore, we evaluated the performance of CineMyoPS with a clinical index, \ie transmurality.

\subsection{Parameter and Ablation Study}
\label{sec:Parameter and Ablation Study}
\subsubsection{Frame Interval Study}

\begin{figure}[htp]
    \centering
    \includegraphics[width=1\linewidth]{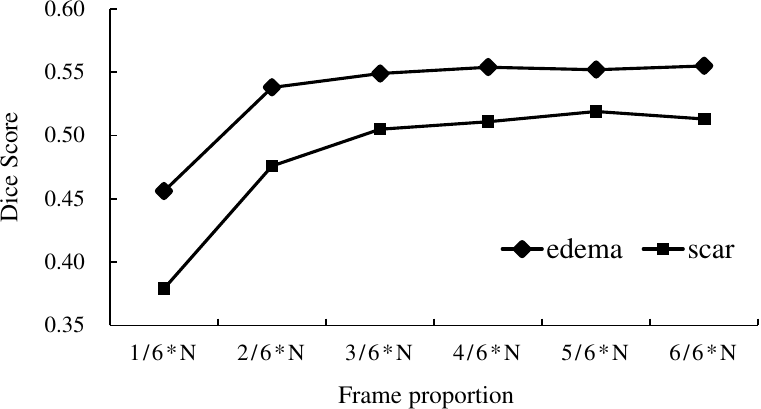}
    \caption{Performance of CineMyoPS with different frame proportions. The average Dice scores of scar and edema were computed on the validation dataset.}
    \label{fig:numberofframe}
\end{figure}

Cine CMR sequences typically comprise 25 to 30 frames per cardiac cycle. We investigated the optimal number of frames required for MyoPS. 
In our experiment, we set the reference image as the starting point and extracted 1/6, 2/6, 3/6, 4/6, 5/6, and 6/6 of the frames within the cardiac cycle. By varying these proportions, CineMyoPS can \hly{aggregate time-series} features relevant to the AAR. \MyFig \ref{fig:numberofframe} illustrates the performance of CineMyoPS on the validation dataset. The results \hly{indicate} that CineMyoPS initially improved with an increasing number of frames and reached a plateau when 4/6  of the total frames were used. Therefore, we adopted 4/6 of the frames from each cine CMR sequence for MyoPS in subsequent experiments, as \hly{this configuration} reached an optimal balance between information richness and computational efficiency.

\subsubsection{Feature Effectiveness Study}

\begin{table*}[ht] 
		\resizebox{1\textwidth}{!}{ 
			\begin{tabular}{llll|llllll|llllll}
				\hline
				\multirow{2}{*}{Methods} & \multicolumn{3}{c}{Features}& \multicolumn{6}{|c}{Scar  }& \multicolumn{6}{|c}{Edema }\\ 
				
				& I  & $\Phi$ & L   & Dice  $\uparrow$  & Pre $\uparrow$ & Sen  $\uparrow$ & Spe  $\uparrow$ & NPV  $\uparrow$ &  HD (mm)   $\downarrow$ & Dice  $\uparrow$  & Pre $\uparrow$ & Sen  $\uparrow$ & Spe  $\uparrow$ & NPV  $\uparrow$ &    HD (mm)   $\downarrow$ \\
				\hline

			  {MyoPS$_\text{I}$}  & \checkmark & $\times$ & $\times$ & 0.41 $\pm$ 0.17 *& 0.66 $\pm$ 0.21 & 0.32 $\pm$ 0.16 * & 0.94 $\pm$ 0.04 & 0.77 $\pm$ 0.08 * & 26.47 $\pm$ 12.77 & 0.50 $\pm$ 0.16 * & 0.76 $\pm$ 0.14 & 0.41 $\pm$ 0.17 *& 0.89 $\pm$ 0.09 & 0.64 $\pm$ 0.12 ** & 28.13 $\pm$ 10.65  \\

                {MyoPS$_\text{L}$} & $\times$& $\times$ & \checkmark & 0.45 $\pm$ 0.17 & 0.71 $\pm$ 0.16 & 0.36 $\pm$ 0.17 & 0.94 $\pm$ 0.04 & 0.78 $\pm$ 0.08 & 25.01 $\pm$ 11.26 & 0.55 $\pm$ 0.12 * & 0.77 $\pm$ 0.12 *& 0.47 $\pm$ 0.14 & 0.88 $\pm$ 0.07 & 0.66 $\pm$ 0.12  ** & 26.84 $\pm$ 12.28  \\
                
			{MyoPS$_{\Phi}$}    & $\times$ & \checkmark & $\times$ & 0.47 $\pm$ 0.15 & 0.72 $\pm$ 0.15 & 0.38 $\pm$ 0.16 & \textbf{0.94 $\pm$ 0.03} & 0.79 $\pm$ 0.07 & 25.05 $\pm$ 10.11 & 0.58 $\pm$ 0.12 & 0.79 $\pm$ 0.10 & 0.50 $\pm$ 0.15 & 0.89 $\pm$ 0.06 & 0.68 $\pm$ 0.12 & 23.85 $\pm$ 10.36  \\
            
                \hdashline
				{MyoPS$_\text{IL}$} & \checkmark & $\times$  & \checkmark & 0.42 $\pm$ 0.15 $\ddagger$ & 0.68 $\pm$ 0.18 & 0.32 $\pm$ 0.14 $\ddagger$ & 0.94 $\pm$ 0.04 & 0.77 $\pm$ 0.08  $\ddagger$ & 27.44 $\pm$ 11.79 $\ddagger$ & 0.50 $\pm$ 0.15 $\ddagger$ & 0.78 $\pm$ 0.15 & 0.41 $\pm$ 0.16 $\ddagger$ & \textbf{0.91 $\pm$ 0.06} & 0.65 $\pm$ 0.12 $\ddagger$ & 25.99 $\pm$ 11.25  \\

    			{MyoPS$_{\text{I} \Phi}$}  & $\times$ & \checkmark & \checkmark & 0.48 $\pm$ 0.16 $\ddagger$ & 0.71 $\pm$ 0.17 & 0.39 $\pm$ 0.16 $\ddagger$ & 0.94 $\pm$ 0.04 & 0.79 $\pm$ 0.07 $\ddagger$  & 24.22 $\pm$ 11.17 $\dagger$ & 0.56 $\pm$ 0.14 $\dagger$ & 0.80 $\pm$ 0.11 & 0.47 $\pm$ 0.17 $\dagger$ & 0.90 $\pm$ 0.06 & 0.67 $\pm$ 0.12 $\dagger$ & 23.72 $\pm$ 10.89  \\

               {MyoPS$_{\Phi \text{L}}$}     & \checkmark & \checkmark & $\times$ & \textbf{0.54 $\pm$ 0.14} & \textbf{0.73 $\pm$ 0.14} & \textbf{0.46 $\pm$ 0.18} & 0.93 $\pm$ 0.05 & \textbf{0.81 $\pm$ 0.07} & \textbf{20.69 $\pm$ 9.13} & \textbf{0.60 $\pm$ 0.13} & \textbf{0.82 $\pm$ 0.08} & \textbf{0.51 $\pm$ 0.16} & 0.90 $\pm$ 0.06 & \textbf{0.69 $\pm$ 0.12} & \textbf{22.68 $\pm$ 8.60}  \\

                % \hdashline
				{{MyoPS$_{\text{I} \Phi \text{L}}$}} & \checkmark & \checkmark & \checkmark & 0.51 $\pm$ 0.13 & 0.72 $\pm$ 0.16 & 0.43 $\pm$ 0.14 & 0.93 $\pm$ 0.04 & 0.80 $\pm$ 0.06 $\dagger$  & 24.34 $\pm$ 10.56 $\dagger$ & 0.56 $\pm$ 0.13 $\ddagger$ & 0.79 $\pm$ 0.11 & 0.48 $\pm$ 0.16 $\dagger$ & 0.90 $\pm$ 0.06 & 0.67 $\pm$ 0.12 $\ddagger$ & 23.38 $\pm$ 10.23  \\

         % {{MyoPS$_{\text{I} \Phi \text{L}}$}} & \checkmark & \checkmark & \checkmark&   &   &   &   &   &   &   &   \\
				  \hline
			\end{tabular}
 		}{
			\caption{Performance of  CineMyoPS \hly{with different feature inputs to the MyoPS module}. The results \hly{were} evaluated on \hly{the} validation dataset. ``$\text{I}$'', ``${\Phi}$'', and ``$\text{L}$'' indicate texture, motion, and anatomy features, respectively.  The best results are highlighted in \textbf{bold}. \hly{Asterisks (*) indicate  significant differences between {MyoPS$_{\Phi}$} and all other methods in the upper panel, while daggers ($\dagger$) indicate differences between {MyoPS$_{\Phi \text{L}}$} and all other methods in the lower panel. The number of  symbols represents the level of significance: $*$ for $p < 0.05$, $**$ for $p < 0.01$, $\dagger$ for $p < 0.05$ and $\ddagger$ for $p < 0.01$.}
            }
			\label{tab:ablation:information}
 		}
\end{table*}

\begin{figure*}[htpb]
    \centering
    \includegraphics[width=1\textwidth]{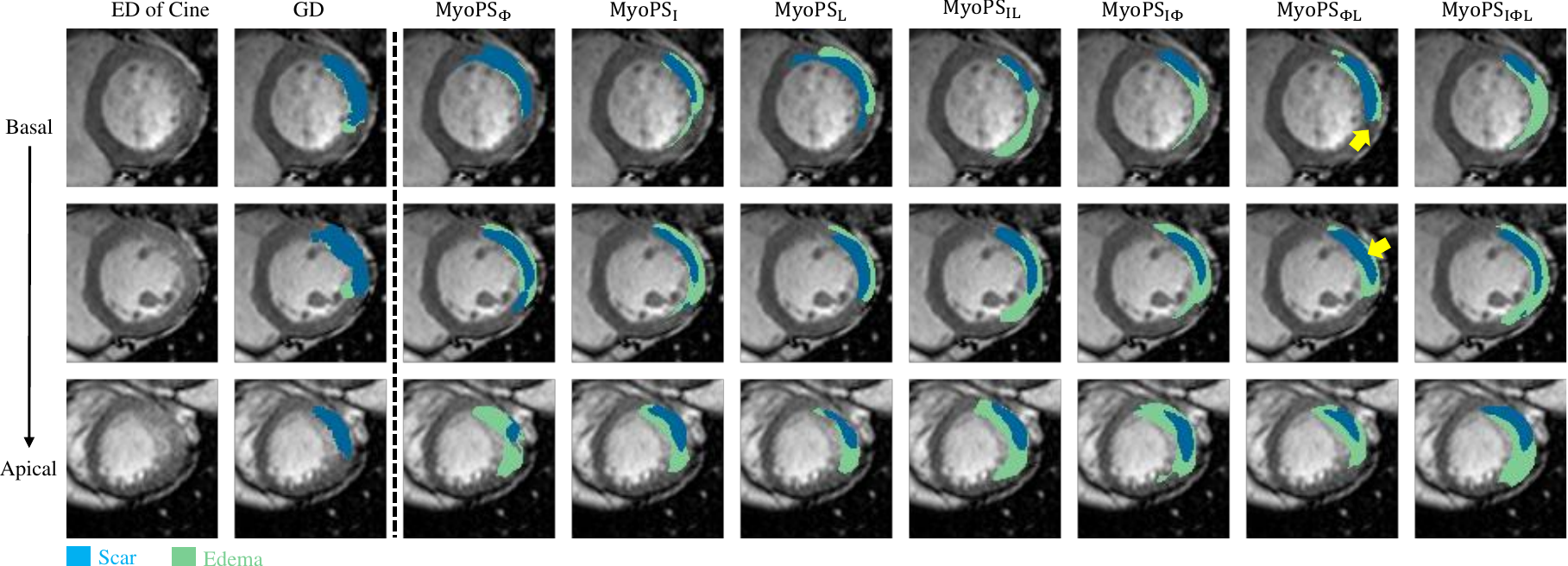}
    \caption{Visualization of MyoPS results by using different features. Yellow arrows indicate the regions where the methods achieved better results than other methods. GD: gold standard.} 
    \label{fig:information}
\end{figure*}

Motion, anatomy, and texture are correlated with MI. We implemented variants of CineMyoPS to verify the effectiveness of different features.  Here, we adjusted \hly{the} MyoPS module of CineMyoPS by combining different features. Note that we transformed texture and anatomy features \hly{using} motion fields before feeding them into \hly{the} MyoPS module (see \MyFig \ref{fig:myops_net}). 
\begin{itemize}
     \item {{MyoPS$_{\text{I} \Phi \text{L}}$}}: \hly{The full version of} CineMyoPS, \hly{where the MyoPS module utilizes all features:} motion ($\Phi$), anatomy ($\text{L}$) and texture ($\text{I}$).
    \item {MyoPS$_{\Phi}$}, {MyoPS$_\text{I}$} and {MyoPS$_\text{L}$}: \hly{Variants of} 
 CineMyoPS \hly{where the MyoPS module utilizes a single type of feature, such as} $\Phi$, $\text{I}$, or $\text{L}$. 
    \item {MyoPS$_{\text{I} \Phi}$}, {MyoPS$_\text{IL}$} and {MyoPS$_{\Phi \text{L}}$}: \hly{Variants of} CineMyoPS \hly{where the MyoPS module} utilizes two types of features. 
\end{itemize}

The upper panel of Table \ref{tab:ablation:information} shows the results of CineMyoPS variants \hly{where the MyoPS module} utilized a single feature. MyoPS$_{\Phi}$ achieved the best Dice, Pre, Sen, Spe, and NPV  among all methods. \hly{For scar segmentation, \hly{MyoPS$_\Phi$} achieved a Dice score that was 0.06 ($p < 0.05$)  higher than  MyoPS$_\text{I}$.}  \hly{For edema segmentation, MyoPS$_\Phi$ achieved Dice scores that were 0.08 ($p < 0.05$) and 0.03 ($p < 0.05$) higher than those of MyoPS$_\text{I}$ and MyoPS$_\text{L}$, respectively.} These results demonstrated that motion is more effective than texture \hly{and anatomy features}.  

The lower panel of Table \ref{tab:ablation:information} shows the results of CineMyoPS where the MyoPS module utilized multiple features. Four combinations of  $\mathbf{I}$, $\mathbf{\Phi}$, and $\mathbf{\hat{A}}$  were evaluated for MyoPS. Among all combinations, {MyoPS$_{\Phi \text{L}}$}  achieved the best performance (\hly{\ie Dice, Pre, Sen, NPV, and HD}) by leveraging motion and anatomy features. \hly{Notably, MyoPS$_{\Phi \text{L}}$ significantly outperformed the other methods in Dice score for edema segmentation ($p < 0.05$).} This finding highlights the effectiveness of integrating motion and anatomy features for MyoPS. Consequently, we adopted motion and anatomy features for MyoPS in \hly{the} following sections.

Interestingly, incorporating more features did not always benefit MyoPS.  
\hly{For example, adding texture features to {MyoPS$_{\Phi \text{L}}$} resulted in a statistically significant reduction of 0.04 ($p < 0.01$) in the Dice score for edema segmentation.} This discrepancy arises because the texture feature may become redundant when anatomy and motion features are already present. Therefore, adding the texture feature may not always improve overall performance.

\MyFig \ref{fig:information} shows a typical case of MyoPS results. In \hly{the} basal and middle slices, MyoPS$_{\Phi \text{L}}$  \hly{produced} visually better results \hly{than the other methods}. This is consistent with the quantitative \hly{results} of Table \ref{tab:ablation:information}. In \hly{the} apical slices, the CineMyoPS variants \hly{tended} to produce false positive edema regions, illustrating the challenges of performing accurate MyoPS in these areas.  

\begin{figure}[htp]
    \centering
    \includegraphics[width=1\textwidth]{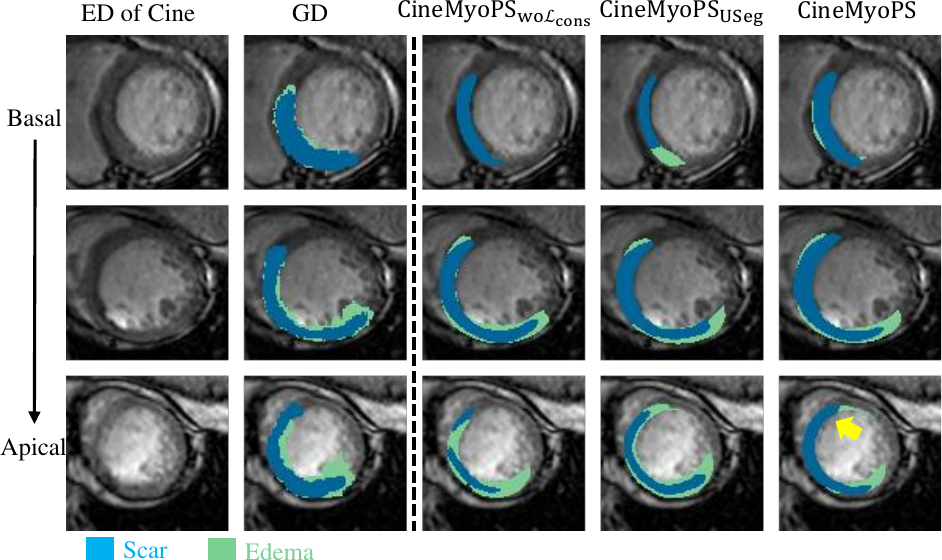}
    \caption{Visualization of myocardial segmentation results \hly{for} variants of CineMyoPS. The yellow arrows mark the areas where CineMyoPS achieved better performance than other variants.}
    \label{fig:ablation_seg}
\end{figure}

\begin{figure}[htpb]
    \centering
    \includegraphics[width=0.9\textwidth]{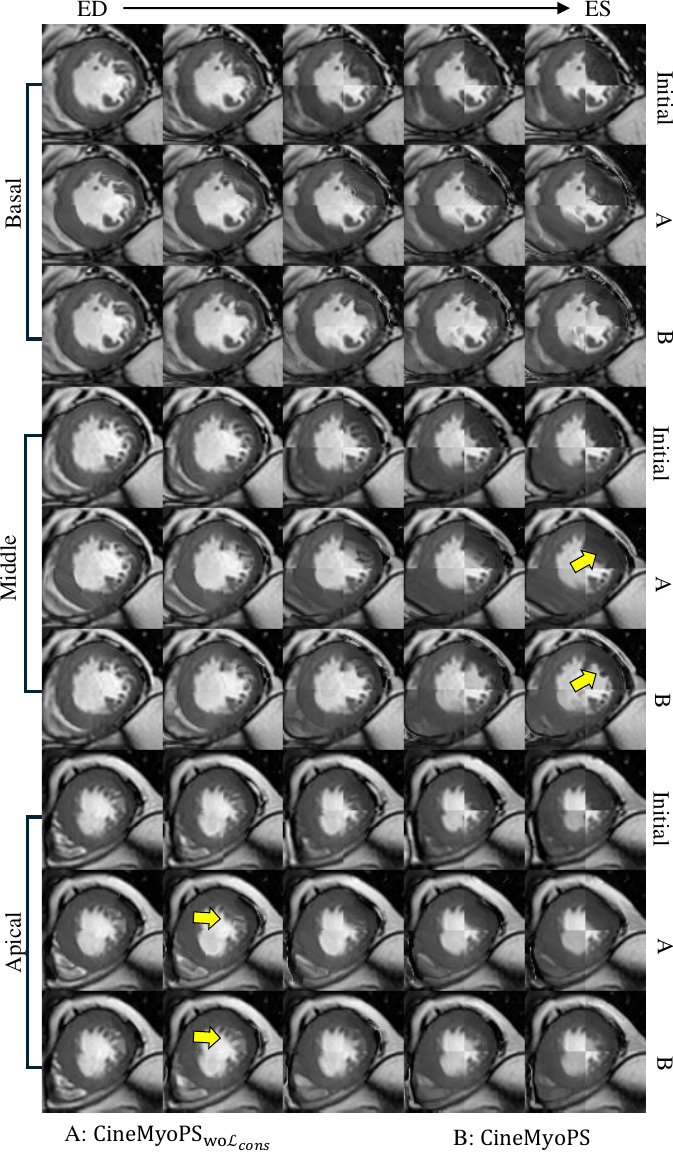}
    \caption{Checkerboard visualization of $I_r$ and $\tilde{I}_i$ frames. The yellow arrows \hly{highlight} the areas where CineMyoPS obtained better results than CineMyoPS$_{\text{wo} \mathcal{L}_{cons}}$.  Please zoom in \hly{to view} the comparison details.}
    \label{fig:motion}
\end{figure}

\begin{figure}[htpb]
    \centering
    \includegraphics[width=1\textwidth]{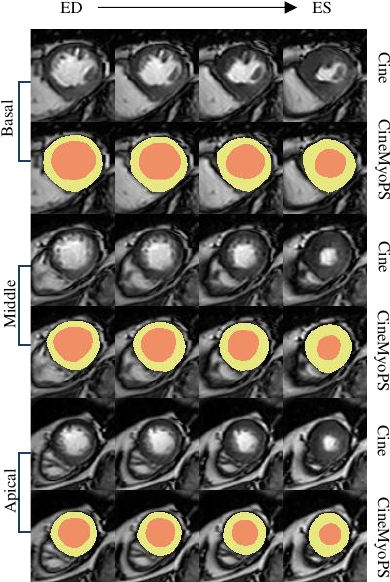}
    \caption{Visualization of myocardial segmentation results from a cine CMR sequence. The segmentation results were \hly{generated by} the anatomy segmentation module of CineMyoPS.}
    \label{fig:myo_seg}
\end{figure}

\begin{table}[htpb]
\resizebox{0.9\textwidth}{!}{
\begin{tabular}{lll}
\hline
{Methods}    &  {Scar  (Dice) $\uparrow$}  & {Edema (Dice)  $\uparrow$}  \\
\hline
{CineMyoPS$_{\text{wo} \mathcal{L}_{cons}}$}  &     0.49 $\pm$ 0.16 ** &   0.57 $\pm$ 0.13    \\
{CineMyoPS$_\text{USeg}$}  & 0.46 $\pm$ 0.15 **  &     0.56 $\pm$ 0.13 *        \\
{CineMyoPS} & \textbf{0.54 $\pm$ 0.14} & \textbf{0.60 $\pm$ 0.13}         \\
\hline
\end{tabular}
}{
\caption{Performance of different variants of CineMyoPS for scar and edema segmentation. \hly{Asterisks (*) indicate statistically significant differences between {CineMyoPS} and all other methods. The number of asterisks represents the level of significance: $*$ for $p < 0.05$ and $**$ for $p < 0.01$.}}
\label{tab:ablation_study}
}
\end{table}

\subsubsection{Consistency Loss and Time-series Aggregate Effectiveness Study}

We implemented two variants of CineMyoPS to investigate the effectiveness of consistency loss and time-series aggregation strategy:
\begin{itemize}
    \item  CineMyoPS$_{\text{wo} \mathcal{L}_{cons}}$:  \hly{A variant of CineMyoPS}  trained without using consistency loss.  
    \item  CineMyoPS$_\text{USeg}$:  \hly{A variant of CineMyoPS} where the MyoPS module was replaced by a \hly{Unet\cite{isensee2021nnu}}, \hly{one of the most straightforward models to perform segmentation in multi-task methods, as demonstrated in \cite{oksuz2020deep}.}
\end{itemize}

\MyTab \ref{tab:ablation_study} lists the Dice scores of variants of CineMyoPS. Without using consistency loss, CineMyoPS suffered performance \hly{degradation} (scar: 0.05, $p<0.01$). This indicates the benefit of \hly{the} consistency loss. Meanwhile, CineMyoPS obtained better results (scar: 0.08, $p<0.01$; edema: 0.04, $p<0.05$) than CineMyoPS$_\text{USeg}$. This reveals that the time-series aggregation strategy can improve \hly{MyoPS}. Moreover, \MyFig \ref{fig:ablation_seg} visualizes a typical case of MyoPS results. CineMyoPS achieved better segmentation details as marked by arrows, which is consistent with the quantitative results in \MyTab \ref{tab:ablation_study}.

\hly{Additionally,} \MyFig \ref{fig:motion} visualizes the checkerboard \hly{comparison of} $I_r$ and transformed $I_i$.  As indicated by yellow arrows, CineMyoPS captured finer details compared to CineMyoPS$_{\text{wo} \mathcal{L}_{cons}}$. \MyFig \ref{fig:myo_seg} shows \hly{the} myocardium segmentation results of a cine CMR sequence.  \hly{It is worth noting that only the reference frame labels were used} to train the anatomy segmentation module of CineMyoPS. Nevertheless, CineMyoPS successfully segmented the anatomy across all frames from end-diastole to end-systole. \hly{These results} demonstrate that the consistency loss enabled CineMyoPS to more accurately extract DDF (motion) and anatomy \hly{features}.

\subsection{Comparison Study}
\label{sec:Performance and Comparisons}
\subsubsection{MyoPS Performance Study}

\begin{table*}[ht] 
		\resizebox{1\textwidth}{!}{ 
			\begin{tabular}{l|llllll|llllll}
				\hline
				\multirow{2}{*}{Methods} & \multicolumn{6}{|c}{Scar  }& \multicolumn{6}{|c}{Edema }\\ 
				
				& Dice $\uparrow$  &Pre $\uparrow$ &  Sen $\uparrow$ &    Spe $\uparrow$ &   NPV $\uparrow$ &   HD (mm)   $\downarrow$ & Dice $\uparrow$  & Pre  $\uparrow$   & Sen $\uparrow$ &  Spe $\uparrow$ &   NPV $\uparrow$ &   HD (mm)   $\downarrow$ \\
				\hline 

			{nnUnet} & 0.42 $\pm$ 0.17 ** & 0.51 $\pm$ 0.21 ** & 0.42 $\pm$ 0.20 ** & \textbf{0.90 $\pm$ 0.04} & 0.85 $\pm$ 0.09 * & 29.68 $\pm$ 11.14 ** & 0.47 $\pm$ 0.14 ** & 0.72 $\pm$ 0.17 ** & 0.42 $\pm$ 0.16 ** & 0.90 $\pm$ 0.06 & 0.68 $\pm$ 0.16 ** & 28.94 $\pm$ 10.50 *  \\
            
				OFSeg & 0.49 $\pm$ 0.15 * & 0.53 $\pm$ 0.19 ** & 0.57 $\pm$ 0.21  & 0.87 $\pm$ 0.06 * & \textbf{0.88 $\pm$ 0.08} & 25.51 $\pm$ 10.58 * & 0.55 $\pm$ 0.12 & 0.72 $\pm$ 0.15 ** & \textbf{0.57 $\pm$ 0.17} * & 0.86 $\pm$ 0.08 ** & \textbf{0.74 $\pm$ 0.15} & 26.01 $\pm$ 11.38  \\
                
    			ConvLSTM & 0.47 $\pm$ 0.14 ** & 0.56 $\pm$ 0.19 ** & 0.52 $\pm$ 0.21 * & 0.88 $\pm$ 0.09 * & 0.87 $\pm$ 0.09 * & 24.03 $\pm$ 9.92 *& 0.56 $\pm$ 0.10 & 0.76 $\pm$ 0.13 ** & 0.54 $\pm$ 0.15 & 0.88 $\pm$ 0.09 ** & 0.73 $\pm$ 0.15 & 25.17 $\pm$ 9.81  \\

	2D+1D Unet  & 0.50 $\pm$ 0.14 & 0.59 $\pm$ 0.17 & 0.54 $\pm$ 0.20 * & 0.90 $\pm$ 0.06 & 0.87 $\pm$ 0.08 & 23.47 $\pm$ 11.74 & 0.56 $\pm$ 0.09 & 0.78 $\pm$ 0.13 & 0.53 $\pm$ 0.15 & 0.90 $\pm$ 0.07 & 0.72 $\pm$ 0.15 & 24.65 $\pm$ 11.19  \\

\hdashline

    CineMyoPS  & \textbf{0.53 $\pm$ 0.12} & \textbf{0.60 $\pm$ 0.18} & \textbf{0.57 $\pm$ 0.19} & 0.90 $\pm$ 0.07 & 0.88 $\pm$ 0.09 & \textbf{21.40 $\pm$ 12.24} & \textbf{0.57 $\pm$ 0.08 }& \textbf{0.79 $\pm$ 0.13} & {0.53 $\pm$ 0.14} & \textbf{0.91 $\pm$ 0.07} & 0.72 $\pm$ 0.14 & \textbf{24.24 $\pm$ 11.71}  \\
   
				  \hline
			\end{tabular}
 		}{
			\caption{Performance of different cine CMR based MyoPS methods on the test dataset.  The best results are highlighted in \textbf{bold}. \hly{Asterisks (*) denote statistically significant differences between {CineMyoPS} and all other methods. The number of asterisks represents the level of significance: $*$ for $p < 0.05$ and $**$ for $p < 0.01$.}}
			\label{tab:compare}
 		}
\end{table*}

\begin{figure*}[ht]
    \centering
    \includegraphics[width=1\textwidth]{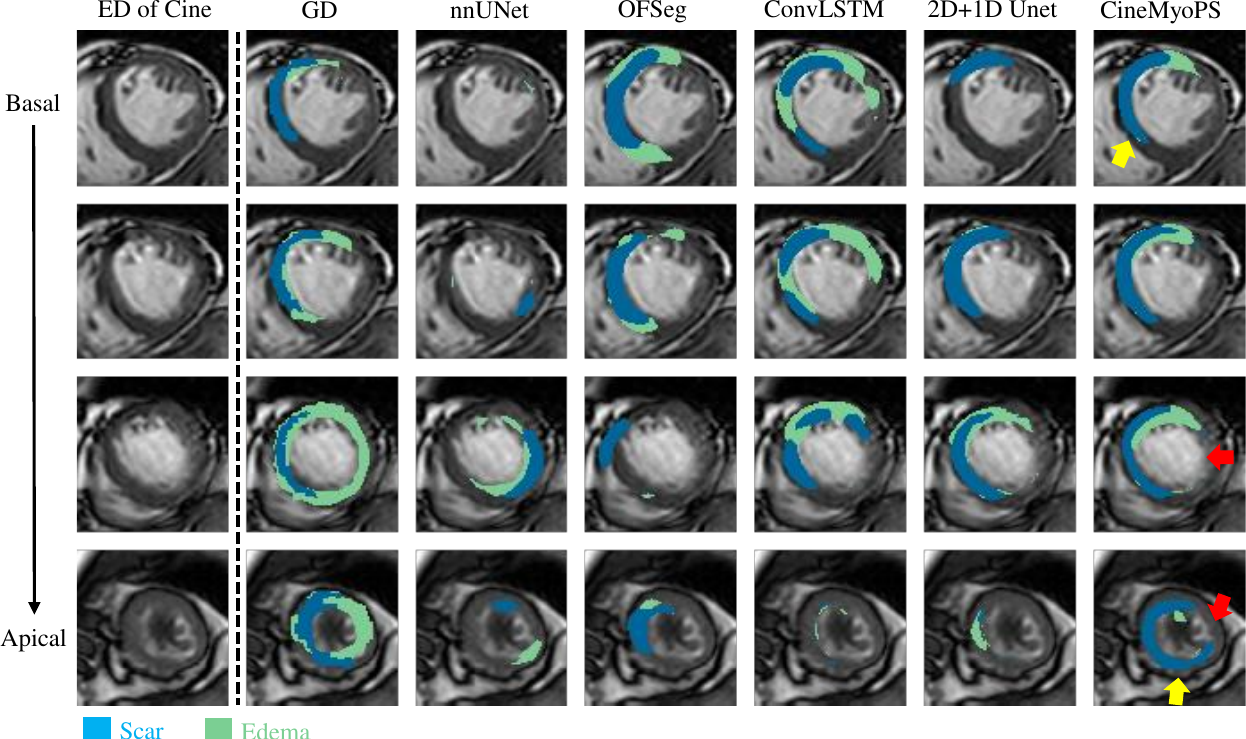}
    \caption{Visualization results of different MyoPS methods. Yellow arrows indicate the regions where CineMyoPS achieved more plausible results than other methods. Red arrows mark the segmentation errors of CineMyoPS.}
    \label{fig:comparsion_method}
\end{figure*}

We implemented four \hly{alternative} MyoPS methods and compared them with the proposed CineMyoPS to assess its effectiveness.
\begin{itemize}
    \item nnUnet: A state-of-the-art segmentation network that directly performs MyoPS from the ED phase of Cine CMR images.
    \item OFSeg:  A segmentation network that predicts myocardial pathologies \hly{based on} myocardial motion \cite{zhang2019deep}. \hly{In this method}, motion features are extracted using an optical flow (OF) method \cite{farneback2003two}. 
    \item ConvLSTM:  A segmentation network \hly{that employs} convolutional LSTM layers to capture spatio-temporal features for MyoPS \cite{shi2015convolutional}.
    \item 2D+1D Unet: A U-shaped segmentation network that predicts myocardial pathologies by integrating 2D-spatial and 1D-temporal features \cite{qiu2017learning}. 
\end{itemize}
\hly{The training loss of CineMyoPS is described by Equation \eqref{eq:totalloss}, while the training loss for the comparison methods is given by Equation \eqref{eq:myops}. Both CineMyoPS and comparison methods were trained with the same strategy, including \hly{the} optimizer and learning rate settings, as outlined in nnUnet \cite{isensee2021nnu}.}

Table \ref{tab:compare} shows the results of different MyoPS methods. CineMyoPS achieved a \hly{0.11 ($p<0.01$) higher Dice score for scar segmentation} compared to nnUnet. Compared to the conventional OFSeg method, CineMyoPS achieved better performance in terms of \hly{Pre and Spe.} Moreover, CineMyoPS estimated motion and performed MyoPS in an end-to-end manner, which simplified the procedure of MyoPS. \hly{Although} 2D+1D Unet and ConvLSTM also extracted spatial features and integrated them along the temporal \hly{axis} for MyoPS, they were less effective compared to CineMyoPS. For example, CineMyoPS achieved \hly{improvements in Sen of}  {0.05} ($p<0.01$) and {0.03} ($p<0.01$)  over ConvLSTM and 2D+1D Unet, respectively, for scar segmentation.  This indicates the effectiveness of the proposed method.

\MyFig \ref{fig:comparsion_method} shows the visual results of different methods. In general, CineMyoPS achieved better visual results than other methods for scar segmentation, as indicated by yellow arrows. However, CineMyoPS may \hly{fail to accurately segment edema}, as indicated by red arrows. This is \hly{because} the motion patterns of scar tissue are more distinguishable from healthy tissue \hly{than} those of edema.

\subsubsection{Related studies from literature}
\label{result_in_literature}

Table \ref{tab:literature} summarizes current MyoPS studies in the literature, including contrast and non-contrast methods. The upper panel lists contrast methods that perform joint scar and edema segmentation. Methods employing contrast-enhanced CMR show promising results, as they are consistent with current clinical settings, \ie with LGE CMR. \hly{Meanwhile,} the performance of contrast methods was better than \hly{that of} CineMyoPS, highlighting the need for enhancements in CineMyoPS to achieve comparable efficacy.

The lower panel lists non-contrast methods that primarily focus on scar segmentation. \hly{Interestingly, some of these} methods not only outperformed CineMyoPS but also exhibited superior results compared to contrast methods. For instance, PSCGAN achieved 0.932 for scar segmentation, while UESTC (a contrast method) and CineMyoPS (our non-contrast method) obtained 0.64 and 0.53, respectively. This variation in performance is \hly{largely} attributed to the differences between datasets. Our dataset presents challenges, as \hly{reflected} by our inter-observer analysis, which yielded average Dice scores of 0.69 for scar segmentation \hly{and 0.73  for edema segmentation}. %\hly{Furthermore, we recruited an additional observer, who achieved Dice scores of 0.73 and 0.75 for scar and edema segmentation, respectively, when delineating myocardial pathologies.  These results highlight the inherent difficulty of the MyoPS task.} 
\hly{Furthermore, we recruited an additional observer to perform MyoPS. The observer achieved Dice scores of 0.73 and 0.75 for scar and edema segmentation, respectively, compared to the gold standard labels. These results highlight the inherent difficulty of the MyoPS task.}

\begin{table}[htpb]
\resizebox{1\textwidth}{!}{
\begin{tabular}{lcc}
\hline
Study    &  Scar {(Dice)}   $\uparrow$ & Edema {(Dice)} $\uparrow$    \\
\hline
UESTC \cite{zhai2020myocardial}, contrast  &  0.64    &   0.70              \\
UMyoPS \cite{ding2023aligning}, contrast  &  0.65   &  0.73           \\
MyoPS-Net \cite{qiu2023myops}, contrast &  0.66  &   0.74             \\
\hdashline
 PSCGAN\cite{xu2020contrast}, non-contrast  &   0.93   &  -             \\
MuTGAN  \cite{xu2018direct}, non-contrast   &    0.90  &  -              \\
MI-Segmentation \cite{zhang2019deep}, non-contrast    &    0.86  &  -               \\
%\hdashline
CineMyoPS, non-contrast &   0.53  &   0.57      \\
\hline
\end{tabular}
}{
\caption{Summary of representative MyoPS studies in \hly{the} literature. \hly{Methods in} the upper panel utilize contrast-enhanced CMR sequences, while \hly{those} in the lower panel rely solely on non-contrast CMR sequences. The performance of reported studies are directly obtained from the original publications.
}
			\label{tab:literature}
}
\end{table}

\subsection{Clinical Quantification Study}
\label{sec:clnical_study}

We further \hly{evaluated} the performance of CineMyoPS using a \hly{clinically relevant} index, \ie \hly{scar transmurality}. To calculate transmurality, we adopted a chord method \cite{sheehan1986advantages}, which divides the myocardium into 100 equally spaced chords. The transmurality of each chord was quantified as (scar pixels / chord pixels)$\times$100\%. Each chord was classified into \hly{one of the} four types according to its transmurality: viable (0 $\sim$ 25\%), likely viable (26\% $\sim$ 50\%), likely nonviable (51\%  $\sim$ 75\%), and nonviable (76\% $\sim$ 100\%) \cite{zhang2022artificial}. 

\begin{figure}[htp]
     \centering
    \makebox[1\textwidth][l]{\ \ \footnotesize {Viable (0\% $\sim$ 25\%) \ \ \ \ \ \ \ \ \ \ \ \ \ \ \ \ \ \ \ \ \ Likely viable  (25\% $\sim$ 50\%)  }}
	\subfigure{\label{fig:zs:25_cinemyops}\includegraphics[width=0.49\textwidth]{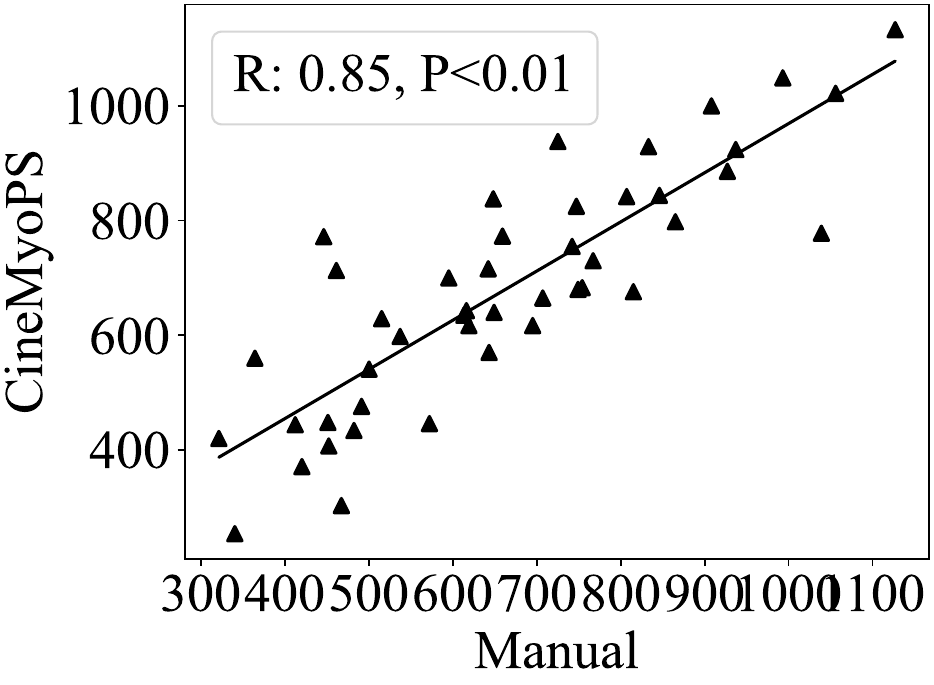}}
	\subfigure{\label{fig:zs:50_cinemyops}\includegraphics[width=0.49\textwidth]{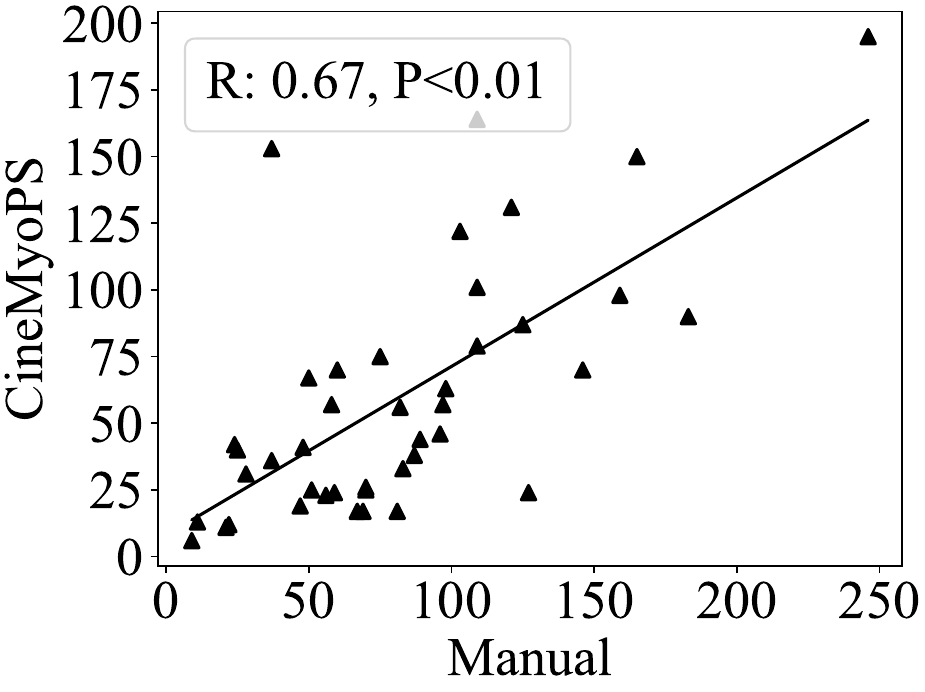}}
	\makebox[1\textwidth][l]{\ \ \ \footnotesize {Likely nonviable (50\% $\sim$ 75\%)  \ \ \ \ \ \ \  Nonviable (75\% $\sim$ 100\%) } }
	\subfigure{\label{fig:zs:75_cinemyops}\includegraphics[width=0.49\textwidth]{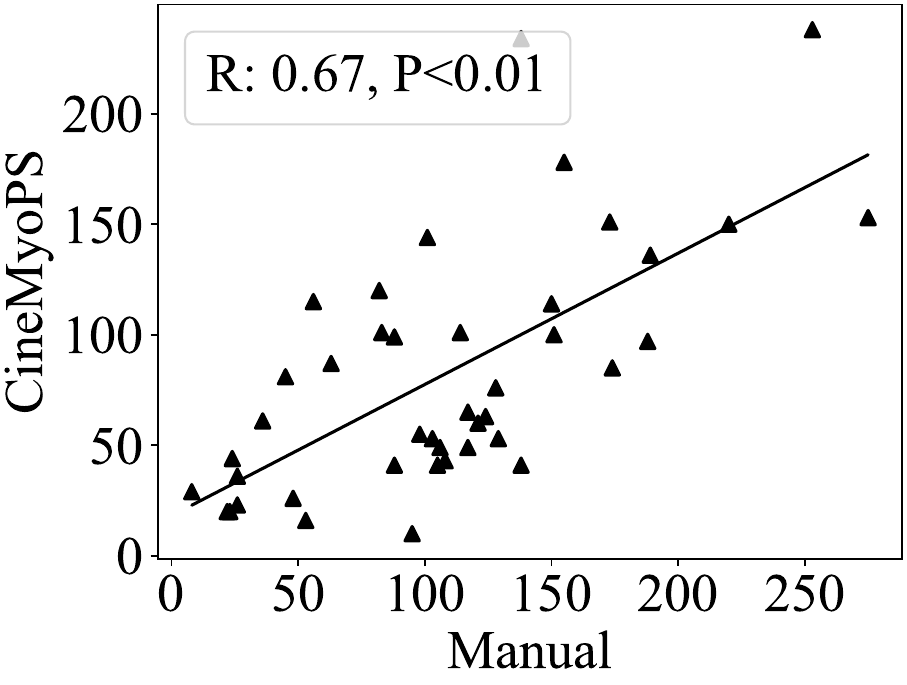}}
	\subfigure{\label{fig:zs:100_cinemyops}\includegraphics[width=0.47\textwidth]{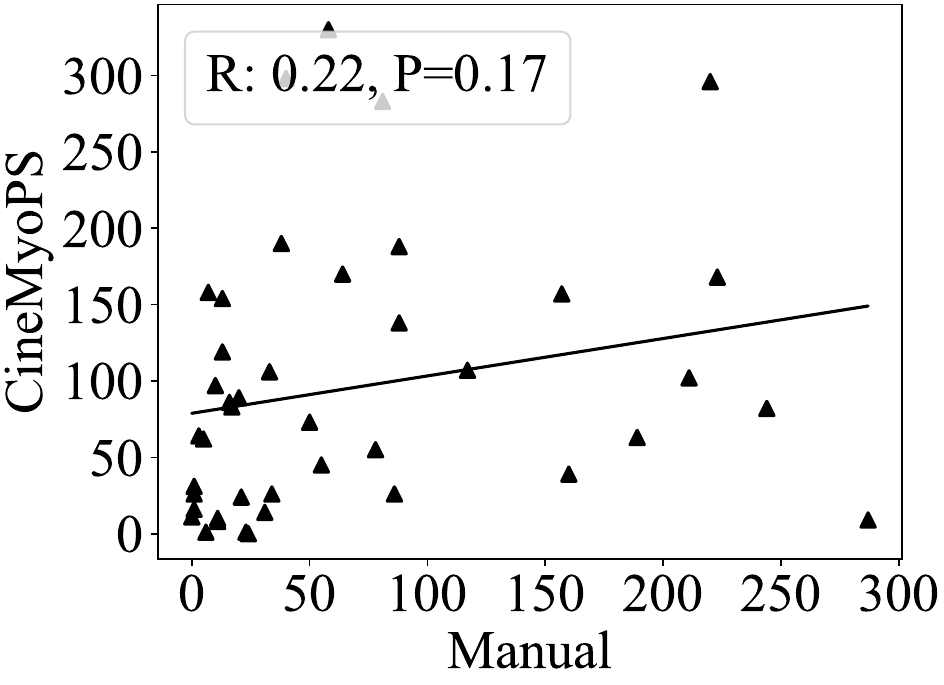}}

	\caption{Assessment of CineMyoPS  for transmurality estimation. We counted the number of chords \hly{within each transmurality category, and analyzed} the correlation between CineMyoPS and manual delineation.  R: Pearson correlation coefficients; P: statistical significance.}

	\label{fig:zs:transmularity} 
\end{figure}

\MyFig \ref{fig:zs:transmularity} shows the scar transmurality assessment results. CineMyoPS achieved a strong correlation \hly{with} manual delineation in viable regions (R=0.85, $p<0.01$). It also demonstrated moderate correlations in likely nonviable (R=0.67, $p<0.01$) and likely viable regions (R=0.67, $p<0.01$). However, CineMyoPS was unsuccessful in estimating scar transmurality in nonviable myocardium (R=0.22, $p=0.17$).  These findings \hly{suggest} that although CineMyoPS is capable of identifying scar regions, its accuracy in delineating the full extent of scar tissue remains limited.

\begin{figure}[bhtp]
    \centering
    \includegraphics[width=1\textwidth]{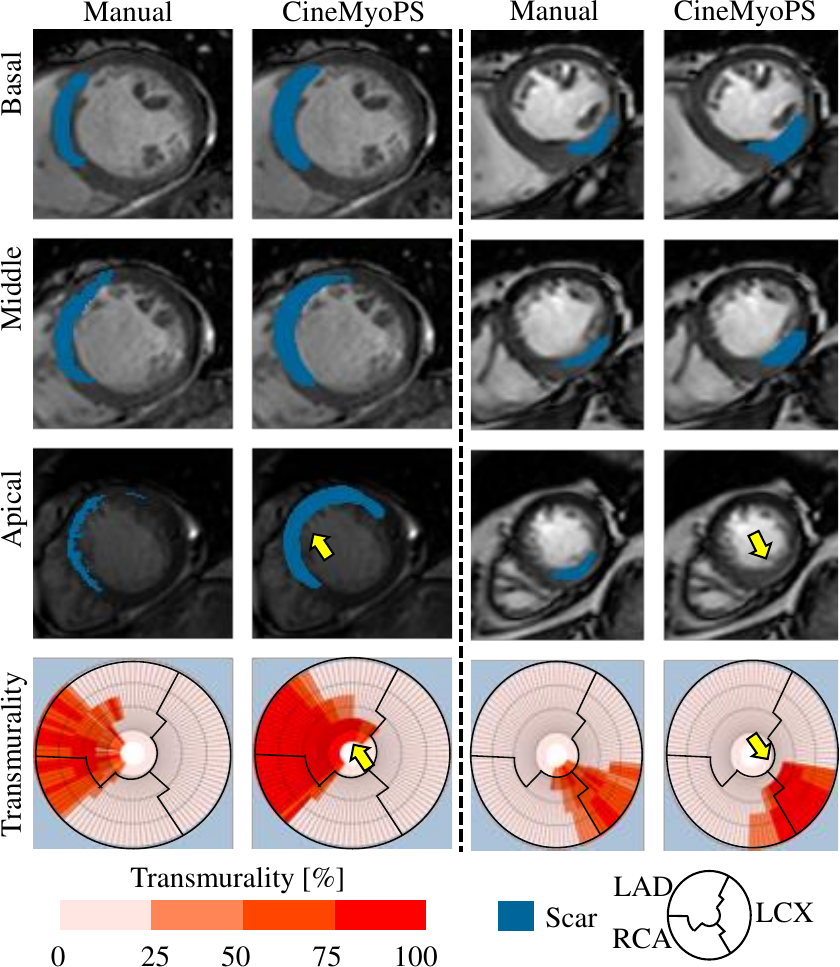}
    \caption{Visualization of two representative scars and their transmurality. Each chord is color-coded according to its transmurality. Yellow arrows indicate areas where CineMyoPS \hly{exhibited inaccurate transmurality estimations.} Bull's-eye plots in the final row \hly{depict} the transmurality of each subject. LAD: left anterior descending artery; LCX: left circumflex artery;  RCA: right coronary artery.}
    \label{fig:vis:transmurality}
\end{figure}

Additionally, \MyFig \ref{fig:vis:transmurality} illustrates the transmurality of two typical scar cases. CineMyoPS achieved comparable visual results \hly{with} manual delineation in basal and middle slices. However,  as indicated by yellow arrows, \hly{its performance declined in the} apical slices, which are commonly considered challenging cases for MyoPS \cite{li2023myops}. Thus, CineMyoPS requires further improvement in apical slices.

\section{Discussion and Conclusion}\label{sec:diss}

This work introduces a novel framework (CineMyoPS) to assess the AAR in MI patients. It explores \hly{the potential of replacing} LGE and T2-weighted images with cine CMR images, \hly{aiming to shorten} the acquisition time of MS-CMR and \hly{eliminate} the injection of contrast agents. Although contrast agents can enhance image quality, they present significant disadvantages, including potential risks of gadolinium deposition and cross-reactivity. CineMyoPS explicitly captures motion, anatomy, and texture features from cine CMR images, allowing for a comprehensive assessment of myocardial pathologies. This not only enhances patient safety but also streamlines the clinical workflow.

In \hly{the} literature, motion has been \hly{widely explored} for scar segmentation \cite{zhang2019deep}. Meanwhile, since MI induces myocardial remodeling, anatomy is also important for MyoPS \cite{o2021automated,avard2022non}. Our experimental results align with these previous studies. \hly{Additionally}, the texture of cine images may exhibit T1/T2 changes for MyoPS \cite{tahir2017acute,jaubert2020free}. \hly{However,} our experimental results suggest that the benefits of including texture features are limited. The upper panel of Table \ref{tab:ablation:information} shows that motion is the most impactful feature, followed by anatomy, with texture being the least significant for MyoPS.

The performance of different cine CMR-based MyoPS methods is influenced by the types of features utilized. This is particularly evident in the performance variability among CineMyoPS variants and competing methods. Competing methods (such as ConvLSTM and 2D+1D Unet) shows minor differences in Table \ref{tab:compare}. The main reason is that the feature extraction capabilities of the competing methods tend to be consistent. Specifically, these competing methods employed learnable techniques such as 2D+1D CNN and ConvLSTM to extract deep latent features for MyoPS. In contrast, the different combinations of explicit motion, anatomy and textural features utilized by CineMyoPS (such as MyoPS$_\text{I}$ and MyoPS$_{\Phi \text{L}}$) lead to larger performance variability, as listed in Table \ref{tab:ablation:information}. Nevertheless, since CineMyoPS allows \hly{the} flexible \hly{integration} of different features, it can effectively mitigate the influence of irrelevant features on MyoPS.

Integrating temporal information \hly{improves the performance of} MyoPS.  Existing methods generally utilize \hly{temporal} information based on spatio-temporal \cite{yan2024coarse,liu2023accurate,qi2024predicting} and 4D \cite{amyar2023gadolinium} CNN techniques. We also observed that integrating time-series features further improves CineMyoPS performance. For instance, without fusing time-series information, CineMyoPS$_\text{USeg}$ \hly{exhibited a decrease in Dice scores}  for scar and edema segmentation, as listed in Table \ref{tab:ablation_study}.  This finding \hly{is consistent with previous studies}. Furthermore, we \hly{investigated the optimal amount of temporal information required} for MyoPS. As shown in Figure \ref{fig:numberofframe}, utilizing 4 out of 6 \hly{frames in a} cardiac cycle is sufficient for MyoPS. 

Another important point to emphasize is that CineMyoPS was trained using pre- and post-contrast cine CMR images, while testing was conducted solely with pre-contrast images. Note that most cine CMR images in clinical practice are acquired after contrast injection, resulting in texture features that exhibit the presence of contrast. As listed in Table \ref{tab:ablation:information}, CineMyoPS achieved Dice scores of 0.41 and 0.50 for the segmentation of scars and edema, respectively, when using texture features. Considering the domain gap \cite{su2023mind} between pre-contrast and post-contrast datasets, we believe that CineMyoPS \hly{can} be improved \hly{by training on a} larger pre-contrast cine CMR dataset.

The performance and interpretability of CineMyoPS require further enhancement for clinical \hly{translation}. A \hly{notable} limitation \hly{is} its inaccuracies in delineating the full extent of MI areas (see \MyFig \ref{fig:zs:transmularity}). Meanwhile, CineMyoPS \hly{encounters} substantial challenges \hly{when processing} apical slices (see \MyFig \ref{fig:information} and \ref{fig:vis:transmurality}), and \hly{may fail to segment} edema, as \hly{shown} in Figure \ref{fig:comparsion_method}. \hly{It is worth noting that MyoPS remains a highly challenging task, as even a human observer was unable to achieve accurate consistency with the gold standard (see Section \ref{result_in_literature})}. Thus, future work \hly{should aim to enhance the performance of} CineMyoPS. \hly{Moreover, although} CineMyoPS can identify the presence of scars and edema, its lack of clinical interpretability may obscure model failures. Future work could \hly{develop synergistic} models for CineMyoPS, paving the way for clinical translation \cite{corral2020digital}.

\bibliographystyle{ieeetr}
\bibliography{refs}
 
\end{document}